# Field- and time-normalization of data with many zeros:
## An empirical analysis using citation and Twitter data[1]

Robin Haunschild[1]   Lutz Bornmann[2]

[1] *r.haunschild@fkf.mpg.de*
Max Planck Institute for Solid State Research, Heisenbergstr. 1, 70569 Stuttgart (Germany)

[2] *bornmann@gv.mpg.de*
Division for Science and Innovation Studies, Administrative Headquarters of the Max Planck Society, Hofgartenstr. 8, 80539 Munich (Germany)

**Abstract**

Thelwall (2017a, 2017b) proposed a new family of field- and time-normalized indicators, which is intended for sparse data. These indicators are based on units of analysis (e.g., institutions) rather than on the paper level. They compare the proportion of mentioned papers (e.g., on Twitter) of a unit with the proportion of mentioned papers in the corresponding fields and publication years. We propose a new indicator (Mantel-Haenszel quotient, MHq) for the indicator family. The MHq is rooted in the Mantel-Haenszel (MH) analysis. This analysis is an established method, which can be used to pool the data from several 2×2 cross tables based on different subgroups. We investigate using citations and assessments by peers whether the indicator family can distinguish between quality levels defined by the assessments of peers. Thus, we test the convergent validity. We find that the MHq is able to distinguish between quality levels in most cases while other indicators of the family are not. Since our study approves the MHq as a convergent valid indicator, we apply the MHq to four different Twitter groups as defined by the company Altmetric. Our results show that there is a weak relationship between the Twitter counts of all four Twitter groups and scientific quality, much weaker than between citations and scientific quality. Therefore, our results discourage the use of Twitter counts in research evaluation.

**Introduction**

Alternative metrics (altmetrics) is a new fast-moving area in scientometrics (Galloway, Pease, & Rauh, 2013). Initially, altmetrics – a collection of many web-based indicators – have been proposed as a supplement to traditional bibliometric indicators. They measure attention related to research papers on internet platforms. The core of altmetrics is gathered from social media platforms, but mentions in mainstream media or in policy documents also belong to the umbrella term altmetrics (National Information Standards Organization, 2016; Work, Haustein, Bowman, & Larivière, 2015). According to Haustein (2016), sources of altmetrics can be grouped into (i) social networks, (ii) social bookmarks and online reference management, (iii) social data (e.g., data sets, software, presentations), (iv) blogs, (v) microblogs, (vi) wikis, and (vii) recommendations, ratings, and reviews.

Recently, some indicators based on altmetrics have been proposed which are normalized with respect to the scientific field and publication year. These indicators were developed because studies have shown that altmetrics are – similar to bibliometric data – field- and time-dependent (see, e.g., Bornmann, 2014). Some fields are more relevant to the general public or a broader audience than other fields (Haustein, Larivière, Thelwall, Amyot, & Peters, 2014). The Mean Normalized Reader Score (MNRS) was introduced by Haunschild and Bornmann (2016) for normalization of data from social bookmarks and online reference management platforms (with a special emphasis on Mendeley readers) (see also Fairclough & Thelwall, 2015). The Mean

Discipline Normalized Reader Score (MDNRS) was tailored specifically to Mendeley by Bornmann and Haunschild (2016b). The MDNRS uses Mendeley disciplines for field normalization. The employed normalization procedures rely on average value calculations across scientific fields and publication years as expected values.

However, normalization procedures based on averages (and percentiles) of individual papers are problematic for data sets with many zeros because averages can get close to zero and only few percentile ranks are occupied (Haunschild, Schier, & Bornmann, 2016). The overview of Work, et al. (2015) on studies investigating the coverage of papers on social media platforms show that less than 5% of the analyzed papers were mentioned on many platforms (e.g., Blogs, or Wikipedia). Erdt, Nagarajan, Sin, and Theng (2016) reported similar findings in their meta-analysis. They found that former empirical studies dealing with the coverage of altmetrics show that about half of the platforms are at or below 5%; except for three (out of eleven) where the coverage is below 10%.

Bornmann and Haunschild (2016a) propose the Twitter Percentile (TP) – a field- and time-normalized indicator for Twitter data. Bornmann and Haunschild (2016a) circumvent the problem of Twitter data with many zeros by including in the TP calculation only journals with at least 80% of the papers having at least 1 tweet each. However, this procedure leads to the exclusion of many journals from the TP procedure.

Very recently, Thelwall (2017a, 2017b) proposed a new family of field- and time-normalized indicators. These indicators are based on units of analysis (e.g., a researcher or institution) rather than on single papers. They compare the proportion of mentioned papers (e.g., on Twitter) of a unit with the proportion of mentioned papers in the corresponding fields and publication years (the expected values). The family consists of the Equalized Mean-based Normalized Proportion Cited (EMNPC) and the Mean-based Normalized Proportion Cited (MNPC). Hitherto, this new family of indicators has only been studied on rather small samples.

In this study, we investigate the new indicator family empirically on a large scale (multiple complete publication years) and add another member to this family. In statistics, the Mantel-Haenszel (MH) analysis is frequently used for pooling the data from multiple 2×2 cross tables based on different subgroups. In this study, we have mentioned and not-mentioned papers of a unit, which have been published in different subject categories and publication years and are compared with the corresponding reference sets. We name the new indicator Mantel-Haenszel quotient (MHq). In the empirical part of this study, we compare the indicator scores with assessments by peers. We are interested whether the indicators can discriminate between different quality levels, which peers assigned to publications. In other words, we investigate the convergent validity of the indicators. The convergent validity can only be tested by using citations, since we can assume that they are related to quality (Diekmann, Naf, & Schubiger, 2012). Thus, the first empirical part is based on citations. In the second part (after confirmation of convergent validity), MHq values are exemplarily presented for Twitter data.

## Indicators for count data with many zeros

The next sections focus on the formulas not only for the calculation of the EMNPC, MNPC, and MHq, but also for the corresponding 95% confidence intervals (CIs). The CI shows the range of possible indicator values: We can be 95% confident that the interval includes the "true" indicator value in the population. Thus, we assume to analyze sample data and infer by using CIs to a larger, inaccessible population (Williams & Bornmann, 2016). Claveau (2016) argues for using inferential statistics with scientometric data as follows: "these observations are realizations of an underlying data generating process … The goal is to learn properties of the data generating process. The set of observations to which we have access, although they are all the actual realizations of the process, do not constitute the set of all possible realizations. In consequence, we face the standard situation of having to infer from an accessible set of observations – what is normally called the sample – to a larger, inaccessible one – the population. Inferential statistics are thus pertinent" (p. 1233).

*Equalized Mean-based Normalized Proportion Cited (EMNPC)*

Thelwall (2017a, 2017b) proposed the EMNPC as an alternative indicator for count data with many zeros. Here, the proportion of mentioned publications is calculated: suppose that the publication set of a group $g$ has $n_{gf}$ papers in the publication year and subject category combination $f$. $s_{gf}$ denotes the number of mentioned papers (e.g., on Twitter). $F$ denotes all publication year and subject category combinations of the publications in the set. The overall proportion of $g$'s mentioned publications is the number of mentioned publications ($s_{gf}$) divided by the total number of publications ($n_{gf}$):

$$p_g = \sum_{f \in F} s_{gf} \Big/ \sum_{f \in F} n_{gf} \qquad (1)$$

However, $p_g$ could lead to misleading results, if the publication set $g$ includes many publications which appeared in fields with many mentioned papers. Thus, Thelwall (2017a, 2017b) proposes to artificially treat $g$ as having the same number of publications in each year and subject category combination. Thelwall (2017a, 2017b) fixes it to the arithmetic mean of numbers in each combination. However, he recommends excluding combinations of $g$ with only a few publications in the analysis. Thus, the equalized sample proportion $\hat{p}_g$ is the average of the proportions in each combination:

$$\hat{p}_g = \frac{\sum_{f \in F} \frac{s_{gf}}{n_{gf}}}{[F]} \qquad (2)$$

The corresponding equalized sample proportion of the world ($w$, i. e., all papers in the analyzed set) is:

$$\hat{p}_w = \frac{\sum_{f \in F} \frac{s_{wf}}{n_{wf}}}{[F]} \qquad (3)$$

In Eqns. (2) and (3), $[F]$ is the number of subject category and publication year combinations in which the group (in case of Eq. (2)) and the world (in case of Eq. (3)) publish. The equalized group sample proportion has the following undesirable property: it treats $g$ as if the average number of mentioned publications does not vary between the subject categories. The ratio of both equalized sample proportions (group and world) is the EMNPC:

$$\text{EMNPC} = \hat{p}_g / \hat{p}_w \qquad (4)$$

According to (Bailey, 1987); Thelwall (2017a), CIs for the EMNPC can be calculated:

$$EMNPC_L = \exp\left(\ln\left(\frac{\hat{p}_g}{\hat{p}_w}\right) - 1.96\sqrt{\frac{(n_g - \hat{p}_g n_g)/(\hat{p}_g n_g)}{n_g} + \frac{(n_w - \hat{p}_w n_w)/(\hat{p}_w n_w)}{n_w}}\right) \quad (5)$$

$$EMNPC_U = \exp\left(\ln\left(\frac{\hat{p}_g}{\hat{p}_w}\right) + 1.96\sqrt{\frac{(n_g - \hat{p}_g n_g)/(\hat{p}_g n_g)}{n_g} + \frac{(n_w - \hat{p}_w n_w)/(\hat{p}_w n_w)}{n_w}}\right) \quad (6)$$

In Eqns. (5) and (6), $n_w$ is the total sample size of the world, and $n_g$ is the total sample size of group $g$. However, according to Thelwall (2017a), these "confidence intervals seem to be only approximations, however, as they can differ from bootstrapping estimates" (p. 133).

*Mean-based Normalized Proportion Cited (MNPC)*

The other indicator proposed by Thelwall (2017a) is the Mean-based Normalized Proportion Cited (MNPC) which is calculated as follows: For each publication which is mentioned at least once (e.g., mentioned on Twitter), the reciprocal of the world proportion mentioned for the corresponding subject category and publication year replaces the number of mentions. All unmentioned publications remain at zero. If $p_{gf} = s_{gf}/n_{gf}$ is the proportion of mentioned publications in set $g$ in the corresponding subject category and publication year combination $f$ and $p_{wf} = s_{wf}/n_{wf}$ the proportion of world's mentioned publications in the same year and subject category combination $f$, then using the number of citations/mentions $c_i$:

$$r_i = \begin{cases} 0, & \text{if } c_i = 0 \\ 1/p_{wf}, & \text{if } c_i > 0 \end{cases} \quad \text{where paper } i \text{ is from year and subject category combination } f$$
$$(7)$$

The MNPC calculation follows the calculation of the MNCS (Waltman, van Eck, van Leeuwen, Visser, & van Raan, 2011) and is defined as:

$$MNPC = \frac{(r_1 + r_2 + \cdots r_{n_g})}{n_g} \quad (8)$$

Thelwall (2016, 2017a) proposes an approximate CI for the MNPC based on a standard formula from Bailey (1987). The lower limit $L$ (MNPC$_{gfL}$) and upper limit $U$ (MNPC$_{gfU}$) for group $g$ in year and subject category combination $f$ are calculated in the first step:

$$MNPC_{gfL} = \exp\left(\ln\left(\frac{\hat{p}_{gf}}{\hat{p}_{wf}}\right) - 1.96\sqrt{\frac{(n_{gf} - \hat{p}_{gf} n_{gf})/(\hat{p}_{gf} n_{gf})}{n_{gf}} + \frac{(n_{wf} - \hat{p}_{wf} n_{wf})/(\hat{p}_{wf} n_{wf})}{n_{wf}}}\right) \quad (9)$$

$$MNPC_{gfU} = \exp\left(\ln\left(\frac{\hat{p}_{gf}}{\hat{p}_{wf}}\right) + 1.96\sqrt{\frac{(n_{gf} - \hat{p}_{gf} n_{gf})/(\hat{p}_{gf} n_{gf})}{n_{gf}} + \frac{(n_{wf} - \hat{p}_{wf} n_{wf})/(\hat{p}_{wf} n_{wf})}{n_{wf}}}\right) \quad (10)$$

The group-specific lower and upper limits are used to calculate the MNPC CIs in a second step:

$$MNPC_L = MNPC - \sum_{f \in F} \frac{n_{gf}}{n_g}\left(\frac{p_{gf}}{p_{wf}} - MNPC_{gfL}\right) \quad (11)$$

$$MNPC_U = MNPC + \sum_{f \in F} \frac{n_{gf}}{n_g}\left(MNPC_{gfU} - \frac{p_{gf}}{p_{wf}}\right) \quad (12)$$

If any of the world proportions are equal to zero, the MNPC cannot be calculated. If any of the group proportions are equal to zero, CIs cannot be calculated. As solutions, either the corresponding subject category and publication year combinations can be removed from the data or a continuity correction of 0.5 can be added to the number of mentioned and not mentioned publications (Thelwall, 2017a). We prefer to use the continuity correction. Plackett (1974) recommends this approach for the calculation of odds ratios. However, according to Thelwall (2017a), these "confidence intervals are only approximations, however, and can differ substantially from bootstrapping estimates" (p. 135).

*Mantel-Haenszel quotient (MHq)*

The recommended method for pooling the data from multiple 2×2 cross tables – based on different subgroups (which are part of a larger population) – is the Mantel Haenszel (MH) analysis (Hollander & Wolfe, 1999; Mantel & Haenszel, 1959; Sheskin, 2007). According to Fleiss, Levin, and Paik (2003) the method "permits one to estimate the assumed common odds ratio and to test whether the overall degree of association is significant. … The fact that the methods use simple, closed-form formulas has much to recommend it" (p. 250). The results of Radhakrishna (1965) point out that the MH approach is empirically and formally valid against the background of clinical trials.

The MH analysis yields a summary odds ratio for multiple 2×2 cross tables. We call this summary odds ratio MHq. If the impact of units in science is compared with reference sets (the world), the 2×2 cross tables (which are pooled) consist of the number of publications mentioned and not mentioned in subject category and publication year combinations $f$ (see Table 1). Publications of group $g$ are part of the publications in the world.

**Table 1. 2×2 subject-specific cross table**

|  | Number of mentioned publications | Number of unmentioned publications |
|---|---|---|
| Group $g$ | $s_{gf}$ | $n_{gf} - s_{gf}$ |
| World | $s_{wf}$ | $n_{wf} - s_{wf}$ |

The MHq calculation starts by defining some auxiliary variables:

$$R_f = \frac{s_{gf}(n_{wf} - s_{wf})}{n_{wf} + n_{gf}} \text{ with } R = \sum_{f=1}^{F} R_f, \tag{13}$$

$$S_f = \frac{(n_{gf} - s_{gf})s_{wf}}{n_{wf} + n_{gf}} \text{ with } S = \sum_{f=1}^{F} S_f, \tag{14}$$

$$P_f = \frac{s_{gf} + (n_{wf} - s_{wf})}{n_{wf} + n_{gf}} \text{ with } Q_f = 1 - P_f \tag{15}$$

The MHq is defined as:

$$\text{MHq} = \frac{R}{S} \tag{16}$$

The MHq CIs are calculated as recommended by Fleiss, et al. (2003). The variance of ln(MHq) is estimated by:

$$\widehat{Var}[\ln(MHq)] = \frac{1}{2}\left\{\frac{\sum_{f=1}^{F}P_f R_f}{R^2} + \frac{\sum_{f=1}^{F}(P_f S_f + Q_f R_f)}{RS} + \frac{\sum_{f=1}^{F}Q_f S_f}{S^2}\right\} \tag{17}$$

Calculation of the MHq CIs is performed as follows:

$$MHq_L = \exp\left[\ln(MHq) - 1.96\sqrt{\widehat{Var}[\ln(MHq)]}\right] \tag{18}$$

$$MHq_U = \exp\left[\ln(MHq) + 1.96\sqrt{\widehat{Var}[\ln(MHq)]}\right] \tag{19}$$

It is an advantage of the MHq that the world average has a value of 1. This is similar to the EMNPC and MNPC and simplifies the interpretation. A further advantage of the MHq is that the result can be interpreted as a percentage relative to the world average. MHq = 1.20, e.g., means that the publication set under study has an impact which is 20% above average.

**Data sets used**

As data sets, we used publications of the Web of Science (WoS) from our in-house database – derived from the Science Citation Index Expanded (SCI-E), Social Sciences Citation Index (SSCI), and Arts and Humanities Citation Index (AHCI) provided by Clarivate Analytics. All papers of the document type "article" with DOI published between 2010 and 2013 were included in the study. The citation counts are restricted to citations with a three-year citation window – excluding the publication year (Glänzel & Schoepflin, 1995). The overlapping WoS subject categories are used for field classification (Rons, 2012, 2014). We include only subject categories in this study where (1) more than 9 papers are assigned to and (2) the number of cited and uncited publications is greater than zero. This should avoid statistical and numerical problems. These restrictions resulted in a dataset including 4,490,998 publications.

We matched the publication data with peers' recommendations from F1000Prime via the DOI. F1000Prime is a post-publication recommendation system of publications which have been published mainly in medical and biological journals. A peer-nominated global "Faculty" of leading scientists and clinicians selects and rates the publications and explains their importance. Thus, only a restricted set from the publications in these disciplines covered is reviewed. Most of the publications are actually not. Faculty members can select and evaluate any publication of interest. Faculty members rate the publications as "Recommended," "Must read", or "Exceptional". This is equivalent to recommendation scores (RSs) of 1, 2, or 3, respectively. Since publications can be recommended multiple times, we calculated an average RS ($\overline{FFa}$):

$$\overline{FFa} = \frac{1}{i_{max}}\sum_{i}^{i_{max}} RS_i \tag{20}$$

The papers are categorized depending on their $\overline{FFa}$ value:

- $\overline{FFa} = 0$: publications which are not recommended (Q0)
- $0 < \overline{FFa} \leq 1.0$: recommended publications with a rather low average score (Q1):
- $\overline{FFa} > 1.0$: recommended publications with a rather high average score (Q2):

We take these three groups (Q0, Q1, and Q2) as our unit of analysis. Following Waltman and Costas (2014), we only considered subject categories where a publication with an F1000Prime recommendation is assigned to.

Whereas the first empirical part of this study is based on citation counts, the second empirical part focusses on Twitter data. The Twitter data were taken from a data set which the company Altmetric has shared with us. We matched the papers with the Twitter information via the paper's DOI. Papers which were unknown to Altmetric were treated as papers with zero tweets. With the number of publications and proportion of uncited and untweeted publications, Table 1 and Table 2 show overviews of the data included in this study. It is clearly visible in Table 2 that Twitter data possess many zeros, but citation data do not.

Table 1. Number of papers included in this study broken down by different sources (citations and Twitter groups), publication year, and $\overline{FFa}$

| Year | $\overline{FFa}$ | Citations | Twitter groups | | | | |
|------|------|-----------|-----|-------------|------------------------|---------------|--------------------|
| | | | all | Researchers | Science communicators | Practitioners | Members of the public |
| 2010 | Q0 | 628,862 | 627,082 | 587,563 | 505,205 | 504,052 | 626,116 |
| 2010 | Q1 | 6,576 | 6630 | 6,286 | 5,941 | 6,189 | 6,622 |
| 2010 | Q2 | 4,368 | 4413 | 4,293 | 4,089 | 4,176 | 4,413 |
| 2011 | Q0 | 681,749 | 683,815 | 669,966 | 614,564 | 598,772 | 683,815 |
| 2011 | Q1 | 6,324 | 6439 | 6,335 | 6,065 | 6,258 | 6,439 |
| 2011 | Q2 | 4,418 | 4494 | 4,459 | 4,391 | 4,424 | 4,494 |
| 2012 | Q0 | 733,813 | 737,074 | 730,917 | 704,025 | 706,521 | 736,992 |
| 2012 | Q1 | 5,826 | 5974 | 5,944 | 5,846 | 5,888 | 5,973 |
| 2012 | Q2 | 5,042 | 5176 | 5,163 | 5,133 | 5,134 | 5,176 |
| 2013 | Q0 | 785,961 | 788,706 | 786,486 | 772,364 | 770,555 | 788,758 |
| 2013 | Q1 | 4,176 | 4254 | 4,249 | 4,208 | 4,233 | 4,255 |
| 2013 | Q2 | 6,361 | 6512 | 6,512 | 6,470 | 6,503 | 6,512 |

Table 2. Proportion of uncited and untweeted papers broken down by different sources (citations and Twitter groups), publication year, and $\overline{FFa}$

| Year | $\overline{FFa}$ | Citations | Twitter groups | | | | |
|------|------|-----------|-----|-------------|----------------------|---------------|----------------------|
|      |      |           | all | Researchers | Science communicators | Practitioners | Members of the public |
| 2010 | Q0 | 10.36 | 95.63 | 98.88 | 99.43 | 99.08 | 96.44 |
| 2010 | Q1 | 0.84 | 86.50 | 95.31 | 98.32 | 97.24 | 88.95 |
| 2010 | Q2 | 0.43 | 76.21 | 89.89 | 96.50 | 94.49 | 80.04 |
| 2011 | Q0 | 10.61 | 87.99 | 97.04 | 98.69 | 98.19 | 89.96 |
| 2011 | Q1 | 1.12 | 69.13 | 89.11 | 95.50 | 93.94 | 72.71 |
| 2011 | Q2 | 0.68 | 51.91 | 76.74 | 90.21 | 89.10 | 56.25 |
| 2012 | Q0 | 10.41 | 72.47 | 92.03 | 96.16 | 95.68 | 77.23 |
| 2012 | Q1 | 1.08 | 38.47 | 75.74 | 84.76 | 84.00 | 44.78 |
| 2012 | Q2 | 0.46 | 23.59 | 56.11 | 72.10 | 75.94 | 29.42 |
| 2013 | Q0 | 10.84 | 68.12 | 89.21 | 94.33 | 93.62 | 73.53 |
| 2013 | Q1 | 1.39 | 31.29 | 70.35 | 81.87 | 77.79 | 37.53 |
| 2013 | Q2 | 0.50 | 21.10 | 50.58 | 69.30 | 69.71 | 26.87 |

In most of the previous studies, which are based on Twitter data, counts of Twitter mentions are used. In this study, we differentiate the Twitter data further on and focus on different Twitter groups: researchers, science communicators, practitioners, and members of the public. These groups are defined by Altmetric based on keyword matching (Adie, 2016):

- Researcher sample keywords: "Post doc", "post-doc", "post-doctoral", ... .
- Science communicator sample keywords: "news service", "Magazine", ... .
- Practitioner sample keywords: "medic", "medical student", "Pediatrician", ... .
- Member of the public is a Twitter user which does not fall into another of the previous three groups.

**Results**

*Empirical analysis using citations*

It is an established way of analyzing the convergent validity of indicators comparing the indicator values with peer evaluations (Garfield, 1979; Kreiman & Maunsell, 2011). Convergent validity is defined as the degree to which two measurements of a construct (here: two indicators of scientific quality) with a theoretical relationship are also empirically related. This approach has been justified by Thelwall (2017b) as follows: "if indicators tend to give scores that agree to a large extent with human judgements then it would be reasonable to replace human judgements with them when a decision is not important enough to justify the time necessary for experts to read the articles in question" (p. 4). Several publications studying the relationship between Research Excellence Framework (REF) outcomes and citations reveal considerable relationships in different fields, such as psychology and biological science (Butler & McAllister, 2011; Mahdi, d'Este, & Neely, 2008; McKay, 2012; Smith & Eysenck, 2002; Traag & Waltman, 2017; Wouters et al., 2015). Similar results have been reported for the Italian research assessment exercise: "The correlation strength between peer assessment and bibliometric indicators is statistically significant, although not perfect. Moreover, the strength of the association varies across disciplines, and it also depends on the discipline internal coverage of the used bibliometric database" (Franceschet & Costantini, 2011, p. 284). Bornmann (2011) shows in an overview of studies on journal peer review that better recommendations from peers are related to higher citation impact of the corresponding papers.

The correlation between citation impact scores and RS from F1000Prime has already been investigated in other studies. The results of Bornmann (2015) reveal that about 40% of papers with RS=1 are highly cited papers; for publications with RS=2 and RS=3 the percentages are 60% and 73%, respectively. Waltman and Costas (2014) report "a clear correlation between F1000 recommendations and citations" (p. 433). The previous results on F1000Prime might point out, therefore, that citation-based indicators differentiate between the three quality levels. Looking at it the other way round, the validity of new indicators does not seem to be given if they do not differentiate.

Against this backdrop, we analyze in this study the ability of MHq, EMNPC, and MNPC to differentiate between the F1000Prime quality groups. Figure 1 shows the MHqs with CIs for Q0, Q1, and Q2 across four publication years.

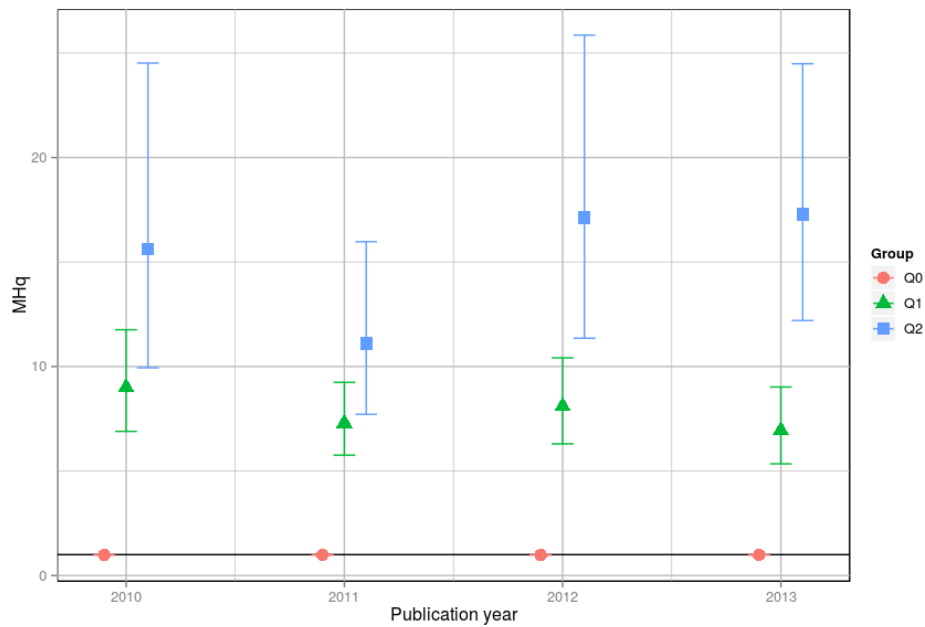

Figure 1. MHqs with CIs for Q0, Q1, and Q2 across four publication years. The horizontal line with MHq=1 is the worldwide average.

The results in Figure 1 point out that the MHq values are very different for Q0, Q1, and Q2. The average MHq for all years is close to (but below) 1 for Q0. The mean MHq for Q1 is about eight times and that for Q2 is about 15 times higher than the mean MHq for Q0. Thus, the MHq indicator seems to separate significantly between Q0, Q1, and Q2; the MHq values seem to be convergent valid with respect to F1000Prime scores.

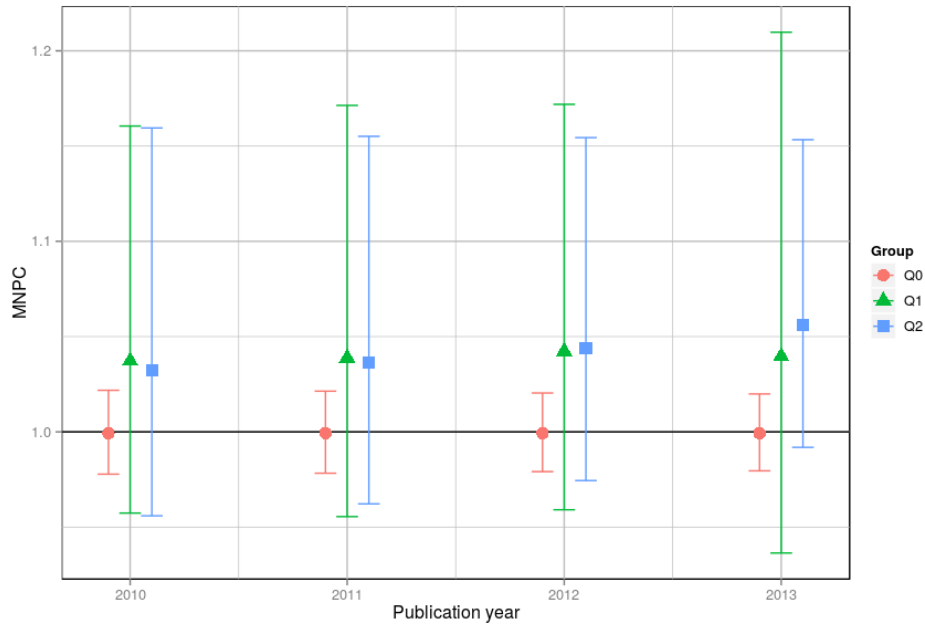

Figure 2. MNPC with CIs for Q0, Q1, and Q2 across four publication years. The horizontal line with MNPC=1 is the worldwide average.

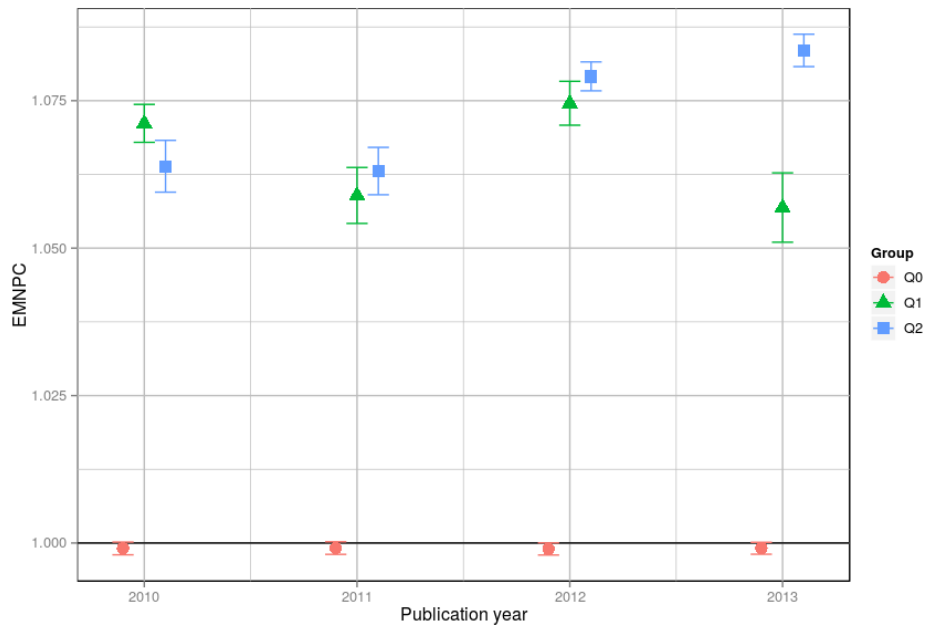

Figure 3. EMNPC with CIs for Q0, Q1, and Q2 across four publication years. The horizontal line with EMNPC=1 is the worldwide average.

The results for MNPC and EMNPC – the two indicators proposed by Thelwall (2017a) – are shown in Figure 2 and Figure 3., It is clearly visible in both figures that the indicator values for all groups are very close to 1. This result does not accord with Figure 1 where the MHq values for Q1 and Q2 are significantly greater than 1. Also, as visible in Figure 2 and Figure 3, the indicator values do not show the expected ordering (analogous to the ordering in Figure 1:

MHq(Q0)<MHq(Q1)<MHq(Q2)) for 2010 in the case of EMNPC and 2010 and 2011 in the case of MNPC. These differences in the results can be interpreted as a first indication that MNPC and EMNPC do not differentiate properly between quality groups as defined by $\overline{FFa}$ values.

*Empirical analysis using Twitter data*

In the previous section, we demonstrated that the MHq is convergent valid using citation data compared with post-publication recommendation scores from F1000Prime. The MHq is able to distinguish between different scientific quality levels as defined by F1000Prime scores. In this section, we start by analyzing the ability of MHq, MNPC, and EMNPC to discriminate between the quality levels Q0, Q1, and Q2 for Twitter data in general. Figure 4 shows the three indicator's performances for Twitter counts. Only the MHq can distinguish between the quality levels for all publication years in the case of Twitter data. The CIs for EMNPC are overlapping for the publication years 2012 and 2013. In the case of MNPC, the CIs are overlapping for all publication years. In general, our observations from citation counts are substantiated by the analysis of Twitter counts.

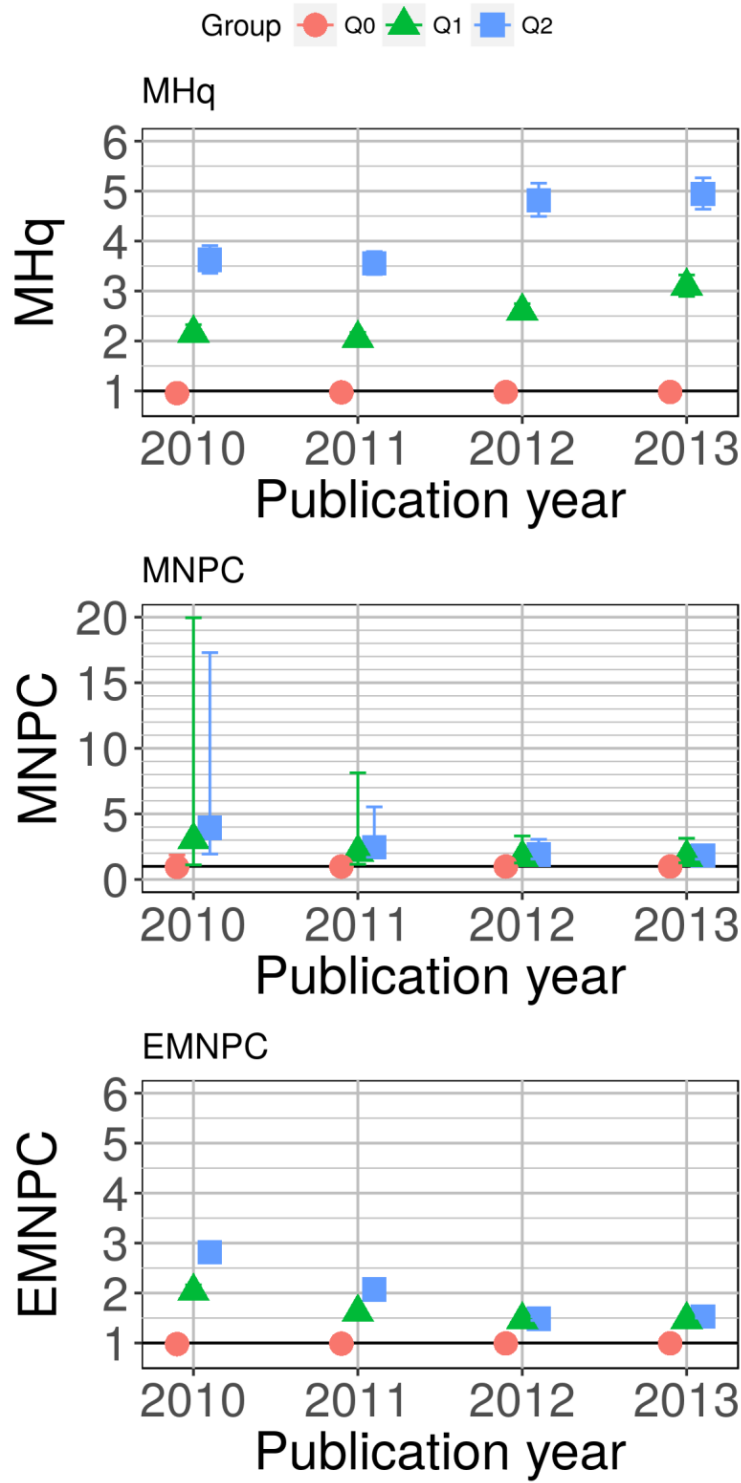

Figure 4. MHq, MNPC, and EMNPC with CIs for Q0, Q1, and Q2 across four publication years for Twitter counts. The horizontal line with indicator values of 1 is the worldwide average.

Next, we determine whether and to which extent different Twitter groups (as defined by the company Altmetric) can distinguish between the same quality levels. Figure 5 shows the MHq results for researchers, science communicators, practitioners, and members of the public.

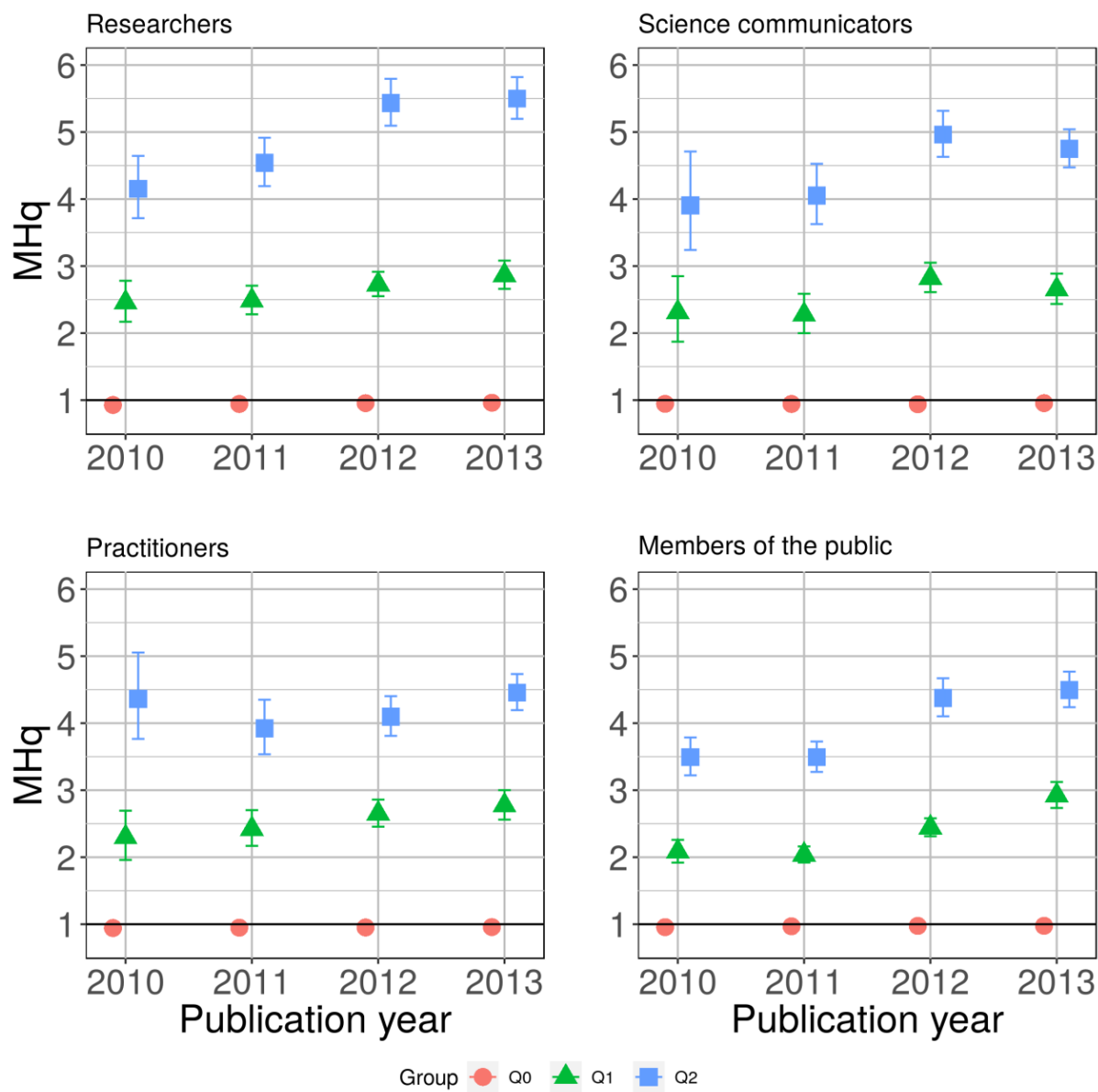

Figure 5. MHq values for Q0, Q1, and Q2 with CIs differentiated by Twitter groups and publication years. The horizontal line with MHq=1 is the worldwide average.

All four Twitter groups show the expected ordering with MHq(Q0) < MHq(Q1) < MHq(Q2). For all four Twitter groups, the quality group Q0 is close to but below 1. The MHq values for the quality groups Q1 and Q2 are between 2 and 4 and between 4 and 6, respectively. Compared to citation data, all four Twitter groups show a much weaker association to scientific quality than citations, by a factor of about three. In Figure 1, which shows the results for citation data, MHq(Q1) is between 7 and 9, and MHq(Q2) is between 11 and 18.

The results in Figure 5 further reveal that the association to scientific quality is on a similar level for all four Twitter groups. Since researchers should assess scientific quality better than the other groups in the figure, the association to quality is somewhat stronger for researchers in the figure than for the other groups. This can be seen from the somewhat higher MHq values for researchers (e. g., 5.5 in 2013) in comparison with the other groups (i. e., 4.5 for practitioners, 4.7 for science communicators, and 4.5 for members of the public in 2013).

**Discussion**

Much of the altmetrics data is sparse (Neylon, 2014). A metric based on many zero values is not informative for research evaluation purposes in the first place (Thelwall, Kousha, Dinsmore, & Dolby, 2016). Thelwall (2017a, 2017b) introduced a new family of field- and time normalized indicators for sparse data including EMNPC and MNPC. Here, the proportion of mentioned publications of a unit (e.g., a researcher) is compared with the expected values (the proportion of mentioned publications in the corresponding publication years and fields).

EMNPC and MNPC differ from most of the indicators used in bibliometrics and altmetrics. Usually, an indicator value is calculated for each publication. The publication-based indicator values can then be aggregated by the user, for example, by averaging or summing. Instead, the indicators of the new family are based on the calculation of the indicator values for publication sets of groups (e.g., universities). This property implies that the new indicators can be used as versatilely as the usual bibliometric (and altmetric) indicators. However, they are able (by construction) to handle data with many zeros properly which the usual indicators are not.

In this study, we added a further variant to the family – the MHq – and analyzed all three variants empirically. We started by analyzing the convergent validity of the indicators based on citation data. Citation data can be used to formulate predictions, which can be empirically validated. Thus, we studied whether EMNPC, MNPC, and MHq are able to validly differentiate between three different quality levels – as defined by RS from F1000 ($\overline{FFa}$). By comparing the indicator values with peer recommendations, we were able to test whether the indicators discriminate between different quality levels. We also analyzed the ability of the three indicators to validly differentiate the quality levels on the basis of Twitter counts.

The results point out that EMNPC and MNPC cannot validly discriminate between different quality levels as defined by peers. The EMNPC and MNPC values are close to the worldwide average – independent of the quality levels. The CIs substantially overlap in many comparisons. Thus, the convergent validity of the EMNPC and MNPC does not seem to be given. In contrast to these indicators, the MHq is able to discriminate between the quality levels.

Because of the positive results for the MHq, we applied the MHq to Twitter counts of four different groups as defined by the company Altmetric: researchers, science communicators, practitioners, and members of the public. If Twitter counts are intended to be used for research evaluation, a substantial relationship to scientific quality should be given. Otherwise, Twitter counts should not be employed in research evaluation. Our investigation of MHq values based on Twitter data reveals a weak relationship between Twitter counts and scientific quality. This relationship is much weaker than that between citation counts and scientific quality.

Our study of the relationship of different Twitter groups' data to scientific quality is directed at specific societal groups. Earlier, we studied the directed societal impact measurement for different status groups of Mendeley data (Bornmann & Haunschild, 2017). Researchers on Twitter show a slightly stronger relationship to scientific quality than other societal groups, but the differences are only minor.

This study follows the important initiative of Thelwall (2017a, 2017b) to develop new indicators for data with many zeros. The current study is the first independent attempt to investigate the new indicator family empirically. This family is important for altmetrics data. Thus, we need further studies focusing on various sources with sparse data (in addition to Twitter). Since F1000 concentrates on biomedicine, future empirical studies should analyze the new family in other disciplines.

The use of F1000Prime peer evaluations as a measure of quality may be a limitation of our study. Peers might be biased from citation counts of the assessed papers or the Twitter accounts they follow. Thus, the recommendation of a paper might be the result of a tweet mentioning the paper. However, F1000Prime faculty members usually recommend papers early after publications so that at least citation rates should not shape their view of many recommended papers.

## Acknowledgements

The bibliometric data used in this paper are from an in-house database developed and maintained by the Max Planck Digital Library (MPDL, Munich, Germany) and derived from the Science Citation Index Expanded (SCI-E), Social Sciences Citation Index (SSCI), and Arts and Humanities Citation Index (AHCI) prepared by Clarivate Analytics (formerly the IP and Science business of Thomson Reuters). The Twitter data were taken from a data set retrieved from Altmetric on June 04, 2016 and stored in a local database and maintained by the Max Planck Institute for Solid State Research (Stuttgart, Germany). The F1000Prime recommendations were taken from a data set retrieved from F1000 in November, 2017. We would like to thank Mike Thelwall for helpful correspondence regarding calculation of the CIs for MNPC and EMNPC. The present study is an extended version of an article (Haunschild & Bornmann, 2017) presented at the 16th International Conference on Scientometrics and Informetrics, Wuhan (China), 16 - 20 October 2017.